\documentclass[12pt]{iopart}

\newcommand{\drt}[2]{\frac{\rmd{#1}}{\rmd{#2}}}

\newcommand{\Reals}{\mathbb{R}}
\newcommand{\cG}{\mathcal G}

\newcommand{\dee}{\mathrm{d}}
\newcommand{\eps}{\varepsilon}

\newcommand{\overbar}[1]{\mkern 1.5mu\overline{\mkern-1.5mu#1\mkern-1.5mu}\mkern 1.5mu}

\usepackage{color}
\usepackage{iopams}
\usepackage{amssymb}
\usepackage{graphicx}
\usepackage[]{algorithm}
\usepackage{algorithmicx}
\usepackage{algpseudocode}
\usepackage{amsthm}

\newcommand{\mi}[1]{{#1}}

\begin{document}

\title[Transform-based filtering for Bayesian inverse problems]{
Transform-based particle filtering for elliptic Bayesian inverse problems}

\author{S Ruchi$^1$, S Dubinkina$^1$ and
M A Iglesias$^2$}
\address{$^1$ Centrum Wiskunde \& Informatica, 
P.O. Box 94079, 1090 GB Amsterdam, The Netherlands}
\address{$^2$ School of Mathematical Sciences, The University of Nottingham, University Park,
Nottingham, NG7 2RD, UK}

\ead{s.dubinkina@cwi.nl}

\vspace{10pt}
\begin{indented}
\item[]August 2017
\end{indented}

\begin{abstract}
We introduce optimal transport based resampling in adaptive SMC. 
We consider elliptic inverse problems of inferring hydraulic conductivity from pressure measurements.
We consider two parametrizations of hydraulic conductivity: by Gaussian random field, and by a set of scalar (non-)Gaussian distributed parameters and Gaussian random fields.
We show that for scalar parameters
optimal transport based SMC performs comparably to monomial based SMC
but for Gaussian high-dimensional random fields optimal transport based SMC
outperforms monomial based SMC.
When comparing to ensemble Kalman inversion with mutation (EKI), 
we observe that for Gaussian random fields, optimal transport based SMC gives comparable or worse performance than EKI depending on the complexity of the parametrization. For non-Gaussian distributed parameters optimal transport based SMC outperforms EKI.
\end{abstract}

\vspace{2pc}




\section{Introduction}

We consider the inverse problem of inferring unknown parameters in models described by partial differential equations (PDEs), given incomplete noisy data/observations of the model outputs. We adopt the Bayesian approach where the unknowns are random functions with a prescribed prior measure that encompasses our prior statistical knowledge of the unknown. The solution to the Bayesian inversion problem is the posterior, i.e. the conditional distribution of the unknown parameters given the observed data. We can use the posterior to compute estimates of the unknown together with the degree of confidence in those estimates. We are interested in problems where the parameter-to-output map from the underlying PDE model is nonlinear. These are particularly challenging problems since the resulting posterior cannot be obtained analytically even when the prior and the noise distributions are assumed Gaussian. Hence, sampling methods are required to approximate (expectations under) the posterior which, in turn, is defined on a very high dimensional space after discretisation of the PDEs that define the forward problem.

Markov chain Monte Carlo (MCMC) is the method of choice to sample the Bayesian posterior \cite{kaipio2005statistical}. In particular, there is a class of MCMC methods constructed in functional settings with mesh-invariant properties suitable for PDE-constrained identification problems~\cite{CRSW13}. However, the most standard version of these methods often exhibit excessively long correlations (e.g. up to $10^4$ \cite{ILS14,KaBeJa14}), a situation particularly exacerbated with highly-peaked (possibly multimodal) posteriors such as those arising when observational noise is small. Very long MCMC long chains (e.g. over $10^7$ steps) are thus required to (i) ensure that MCMC fully explores the posterior measure thus capturing possibly multiple modes and (ii) produce sufficient independent samples to compute accurate posterior statistics. Since every step of MCMC involves at least one PDE solve, these methods become impractical for costly large-scale simulations. While more efficient MCMC can be used to approximate the posterior \cite{0266-5611-30-11-114014,stochasticNewton}, their proposals often required high-order derivatives of the likelihood which are not available in many applications where the simulator is accessible only in a black-box fashion.

Sequential Monte Carlo (SMC) samplers \cite{DelMoral06} offer a different sampling approach for approximating the Bayesian posterior. In the context of large-scale Bayesian inversion, adaptive SMC methods construct particle approximations of a sequence of intermediate measures that interpolate (e.g. via tempering) between the prior and the posterior. Particles and their weights are adapted on-the-fly to enable a controlled transition between those intermediate measures, thus facilitating to gradually move from a simple prior to a possibly complex posterior. The transition between two intermediate measures involves an importance resampling (IR) step by which the particles are weighted according to the tempered likelihood and then resampled according to those weights.  This step is then followed by mutation of particles induced by sampling from a kernel with the IR measure as its invariant measure; this is typically conducted via running MCMC chains with the aforementioned target measure. 

Adaptive SMC samplers for solving Bayesian inverse problems have been proposed in \cite{KaBeJa14} and applied for the identification of the initial condition in the Navier-Stokes equations. This work showed that SMC can produce accurate approximations of the Bayesian posterior at a computational cost an oder of magnitude smaller than those obtained via state-of-the-art MCMC. The same adaptive SMC sampler was used  in \cite{RTM} to infer permeability in a moving boundary problem arising in porous media flow. A theoretical framework for adaptive SMC framework was developed in \cite{BJMS15} and tested numerically by inferring hydraulic conductivity in a groundwater flow model. 

Despite of the computational advantages of using SMC samplers, their computational cost still poses severe limitations for its application to practical large-scale inverse problems. The cost of a single iteration (IR+mutation) within SMC is $J\times N_{\mu}$ where $J$ is the number of particles and $ N_{\mu}$ is the number of mutation MCMC moves. Therefore, each iteration could involve over $10^4$ PDE solves even for relatively small $J$ and $ N_{\mu}$ (i.e. $J=10^3$ and $ N_{\mu}=10$). Hence, if the posterior is complex hence requiring several intermediate measures, the cost of SMC is prohibited unless high performance (HPC) resources are available to scale the cost of SMC with respect to $J$. While parallelisation is indeed one of the main advantages of SMC, the availability of HPC with $10^4-10^5$ processors for typical engineering and geophysical (practical) applications is the exception rather than norm. It is worth mentioning that reducing the cost of SMC via using small number of samples and/or reducing the number of mutation steps can be substantially detrimental to the accuracy of the particle approximation provided by SMC; see for example the work of \cite{RTM} where SMC with limited number of particles ($10^2-10^3$) results in very poor approximations of the Bayesian posterior. Recent work aimed at reducing the computational cost of SMC samplers includes the development of multilevel versions \cite{multilevel,Multilevel_SMC}.

\mi{
\subsection{Contribution of this work}}

Our aim is to investigate the feasibility of an alternative, potentially more computationally affordable, approach to approximate the Bayesian posterior within the adaptive tempering SMC setting for Bayesian PDE-constrained inverse problems \cite{KaBeJa14,BJMS15}.  The proposed approach consist of replacing the resampling step in SMC with a deterministic linear transformation that maps the system of particles that approximate two consecutive measures. At each iteration step within SMC, the transformation is obtained via solving an optimal transportation problem which, in turn, defines a deterministic coupling between two discrete random variables with realisations defined by the particles and with probabilities determined by their corresponding weights. Replacing resampling by an optimal transformation within Bayesian algorithms was proposed in \cite{Re13} where it was shown that the linear transport map leads to samples that converge to the posterior measures in large ensemble limit. In the context of data assimilation of partially observed dynamic systems,  the idea of replacing IR by optimal transport maps is at the core of the so-called ensemble Transform Particle filter (ETPF) \cite{Re13,ReCo15}. \mi{The novelty of our approach lies in transfering the application of optimal transport to compute the transition between measures in the tempering scheme within SMC.}

Numerous work on data assimilation has shown that, when relatively small number of particles are used, ETPF provides more accurate state estimations compared to standard IR-based particles filters due to the sampling errors introduced by resampling. While methods such as ensemble Kalman filter (EnKF) can work well for small ensemble sizes compared to IR-based methods, they rely on Gaussian approximations which is often a severe limitation when the underlying distribution is, for example, multimodal. In contrast, the optimal transport within ETPF does not rely on Gaussian approximations and has been shown to be 1st order consistent for the mean, and to converge to the posterior measure in the large-ensemble size limit \cite{Re13}. Here we investigate whether those well known advantages of ETPF can be exploited  within the setting of adaptive SMC for Bayesian inversion. As a proof-of-concept we apply the proposed algorithm to a Bayesian elliptic inverse problem arising in groundwater flow. The goal is to infer hydraulic conductivity from pressure measurements. We consider two parameterisations of the conductivity field aimed at assessing the method under two levels of complexity. In the first one we assume that the log-conductivity is a smooth function characterised by Gaussian random field under the prior. The second parameterisation consist of a channelised permeability that is described by a set of geometric parameters together with two random fields in the regions inside and outside the channel. While the first parameterisation yields posteriors which are relatively well approximated by Gaussians, the second parameterisation can result in multimodal distributions which are more difficult to capture with Gaussian approximations. 


We compared the performance of the proposed technique against a fully resolved posterior computed by the preconditioned Crank-Nicolson (pcn)-MCMC with sufficient steps to ensure that a chain is properly converged. We then compare the proposed technique against monomial based SMC as well as an ensemble Kalman inversion (EKI) technique that arises naturally from the adaptive SMC setting. This EKI methodology has been proposed in \cite{Chetal18} as an alternative of \cite{EnsembleYO}. Here this approach is modified to incorporate a mutation with the invariant measure.


\section{Forward and Inverse Problem}\label{For_Inv}
Since we consider Bayesian inversion, it demands formulation of both a forward problem and an inverse problem.
The forward problem consists of finding pressure from hydraulic conductivity.
The "inverse" problem consists of two parts. First part is parametrization of hydraulic conductivity by a random variable.
Second part is employment of the Bayes' rule to obtain the posterior distribution of the random variable from a given prior 
and a likelihood. The likelihood involves forward problem evaluation. Thus the Bayesian inversion employs the forward problem within the inverse problem. 
\subsection{Forward Model}\label{forward}
The forward problem consist of the identification of the hydraulic conductivity, $\kappa(x)$, of a two-dimensional confined aquifer for which the physical domain is $D=[0,6]\times [0,6]$. Assuming that the flow within the aquifer is single-phase steady-state Darcy flow, the piezometric head $h(x)$, is given by the solution of \cite{Bear}
\begin{eqnarray}\label{eq1}
-\nabla\cdot \kappa \nabla h&=f &\qquad\textrm{in}~~D
\end{eqnarray}
where $f$ represents recharge term. We use the Benchmark from \cite{Carrera,Hanke,EnsembleYO} where $f$ has the following form
\begin{eqnarray}\label{eq2}
f(x_{1},x_{2})=\left\{\begin{array}{ccc}
0 &\textrm{if}& 0< x_{2}\leq 4,\\
137&\textrm{if}& 4< x_{2}< 5,\\
274&\textrm{if}& 5\leq x_{2} < 6.\end{array}\right.
\end{eqnarray}
and where the boundary conditions are given by
\begin{equation}
\label{eq3}
 h(x_1,0)=100, \quad \frac{\partial h}{\partial x}(6,x_2)=0,\quad
 -\kappa\frac{\partial h}{\partial x}(0,x_2)=500,\quad   \frac{\partial h}{\partial y}(x_1,6)=0.
\end{equation}
We wish to infer $\kappa\in X:=\{f\in L^{\infty}(D;\Reals)|\textrm{ess}\inf_{x \in D} f(x)>0\}$ from point observations of $h$ collected at $M$ locations denoted by $\{x_i\}_{i=1}^{M}\subseteq D$. To this end, we consider smoothed point observations defined by
\[
\ell_j(h) = \int_D \frac{1}{2\pi\eps^2}e^{-\frac{1}{2\eps^2}(x-x_i)^2}h(x)\,\dee x
\]
where $\eps > 0$. Let us define the forward map $G:X\rightarrow \mathbb{R}^{M}$ by
\begin{equation}
\label{eq4}
G(\kappa)=(\ell_1(h),\ldots,\ell_{M}(h)).
\end{equation}
which maps permeability into predictions of hydraulic head at measurement locations. Assume that we have noisy measurements of $\{\ell_j(h)\}_{j=1}^{M}$ of the form 
\begin{equation*}\label{eq5}
y_j=l_j(h)+\eta_j, \qquad j=1,\dots,M
\end{equation*}
where $\eta_j$ represents measurement noise. Our aim is to reconstruct $\kappa\in X$ given $y=(y_{1},\dots,y_M)\in \mathbb{R}^{M}$.

\subsubsection{Parameterisation of permeability}
\label{perm}

\begin{figure}[!ht]
  \centering
  \includegraphics[width=0.5\textwidth]{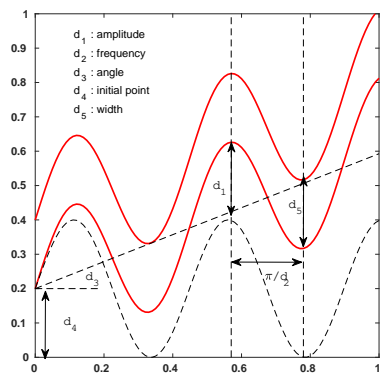}
  \caption{Geometrical configuration of channel flow.\label{fig:ChanGeom}}
  \centering
\end{figure}

We consider the following two parameterisations of the permeability function $\kappa(x)$ that we wish to identify from observations of the Darcy flow model (\ref{eq1})-(\ref{eq3}). 
\begin{itemize}
\item[P1:] For the first model the parameter that we consider is simply the natural logarithm of $\kappa$, i.e. $u(x)=\log{\kappa(x)}$. 

\item[P2:] The second model consist of parameterisation of a piecewise continuous permeability of the form
\begin{equation*}\label{eq5B}
\kappa(x)=\exp(u_{1}(x)) \chi_{D_{c}}(x)+\exp(u_{2}(x))\chi_{D\setminus D_{c}}(x)
\end{equation*}
where $\kappa_{1}=\exp(u_{1}(x))$ and $\kappa_{2}=\exp(u_{2}(x))$ are continuous permeabilities inside and outside a sinusoidal channel with domain denoted by $D_{c}$. The geometry of the channel is parameterized by five parameters $\{d_i \}_{i=1}^5$ as described in Figure~\ref{fig:ChanGeom}. The lower boundary of the channel is given by 
$$x_{2}=d_1\sin(d_2x_{1}/6)+\tan(d_3)x_{1}+d_4$$
where we use the notation $x=(x_{1},x_{2})\in D$ in terms of the horizontal and vertical components. The upper boundary of the channel is given by $x_{2}+d_{5}$. For this permeability model the parameters of interest are comprised in 
\begin{eqnarray*}\label{eq6}
u=  (d_1,\cdots,d_5, u_{1},u_{2})
\end{eqnarray*}
where we assume that each $d_{i}$ is restricted to an interval $A_{i}\equiv [{d_i^-},d_i^+]$. 
\end{itemize}
We define the following parameter space
\begin{equation*}\label{eq7}
U= \left\{\begin{array}{cc}
L^{\infty}(D;\Reals)&\textrm{for P1},\\
\prod_{i=1}^5A_{i}\times L^{\infty}(D;\Reals^2)&\textrm{for P2},\end{array}\right.
\end{equation*}
with metric 
\begin{equation*}\label{eq8}
\vert u \vert_{U}= \left\{\begin{array}{cc}
\vert\vert u \vert \vert_{\infty}&\textrm{for P1},\\
\sum_{i=1}^{5}\vert d_{i}\vert +\vert\vert u_{1} \vert \vert_{\infty}+\vert\vert u_{2} \vert \vert_{\infty}&\textrm{for P2},\end{array}\right.
\end{equation*}

The parameterizations described earlier define an abstract map $F:U\rightarrow X$ from the space of parameter to the space of admissible permeabilities, via
\begin{equation}\label{eq9}
\label{eq:F}
F(u)=\kappa.
\end{equation}
We define the parameter-to-observations map $\cG: U\rightarrow \mathbb{R}^{M}$ by $\cG= G\circ F$ and reformulate the inverse problem (\ref{eq4}) in terms of finding the parameter $u\in U$, given $y\in \mathbb{R}^{M}$ that satisfies
\begin{eqnarray}\label{eq10}
y=\cG(u)+\eta
\end{eqnarray}
for $\eta=(\eta_{1},\dots,\eta_{M})\in \mathbb{R}^{M}$. 
The continuity of the parameter-to-observations map $\cG$ for this, and more general cases, has been established in \cite{Andrew,ILS14}.

%
%
\subsection{The Bayesian Inverse Problem}\label{Bayes}

In order to address the inverse problem formulated via (\ref{eq10}) we adop the Bayesian framework \cite{Andrew} where $\eta$ is a random vector and $u$ is a random function. We put a prior, $\mu_{0}(u)$,  on the unknown $u$, and define the random variable $y\vert u$ under the standard assumption that $\eta\sim N(0,\sigma^2 I)$ independent of $u$. The solution to the inverse problem in the Bayesian setting is the posterior measure on $u\vert y$. In the following sections we introduce the prior and likelihood which by  the infinite-dimensional framework of \cite{Andrew} ensure that the posterior measure exists and is continuous with respect to appropriate metrics.  

\subsubsection{The Prior}\label{prior}

For P1 we consider Gaussian prior $\mu_{0}=N(m,C)$ with mean $m$ and covariance $C$. We define $C$ via a correlation function given by the Wittle-Matern correlation function defined by \cite{matern1}:
  \begin{equation}\label{eq10B}
c(x,y) =\sigma_{0}^{2}\frac{2^{1-\nu}}{\Gamma(\nu)}\Bigg(\frac{\vert x-y\vert }{l} \Bigg)^{\nu}K_{\nu}\Bigg(\frac{\vert x-y\vert }{l}\Bigg),
\end{equation}
where $\Gamma$ is the gamma function, $l$ is the characteristic length scale, $\sigma_{0}^2$ is an amplitude scale and $K_{\nu}$ is the modified Bessel function of the second kind of order $\nu$. The parameter $\nu$ controls the regularity of the samples.

For P2 we assume independence between geometric parameters and log-permeabilities and thus consider a prior of the form
\begin{equation}\label{eq11}
\fl \mu_{0}(du)=  \Pi_{i=1}^{5}\pi_0^{A_{i}}(d_{i})\otimes N(m_1,C_{1}) N(m_2,C_{2})
\end{equation}
where $\pi_0^{A}(x)$ is the uniform density defined by
\begin{equation}\label{pi0}
  \pi_0^{A}(x)=\left\{
  \begin{array}{cc}
  \frac{1}{|A|} &  x \in A,\\
    0 & x \notin A.
  \end{array}
  \right.
\end{equation}
In expression (\ref{eq11}) $N(m_1,C_{1})$ and $N(m_2,C_{2})$ are two Gaussians such as those described earlier in terms of the correlations function from (\ref{eq10B}). 

\subsubsection{The likelihood}
We assume the unknown $u$ is independent of the observational noise $\eta\sim N(0,\sigma^2)$. We note that $y\vert u\sim N(\cG(u),\sigma^2I)$, hence the likelihood is given by 
\begin{eqnarray}\label{eq12}
l(u,y)\propto \exp(-\Phi(u,y))
\end{eqnarray}
where $\Phi(u,y)$ is the data misfit defined by
\begin{eqnarray}\label{eq13}
\Phi(u,y)=\frac{1}{2\sigma^2}\vert\vert y-\cG(u)\vert \vert^2
\end{eqnarray}

\subsubsection{The Posterior}\label{posterior}

The selection of prior measures from subsection \ref{prior} satisfies that $\mu_{0}(U)=1$; i.e. samples from $\mu_{0}$ are in $U$ almost surely \cite{Andrew,ILS14}. This property, together with the continuity of the forward map defined in subsection \ref{forward}, can be used in the Bayesian framework of \cite{Andrew, ILS14} to conclude that (i) the posterior measure $\mu(u)$ on $u\vert y$ exists and is absolutely continuous with respect to the prior; and(ii) $\mu_{0}$ and has a density with respect to $\mu_{0}$ given by the following Bayes' rule 
\begin{eqnarray}\label{eq14}
 \drt{\mu}{\mu_0} = \frac{1}{Z} l(u,y)
\end{eqnarray}
 where
\begin{eqnarray}\label{eq15}
Z=\int_{U} l(u,y)\mu_{0}(\rmd u)
\end{eqnarray}

\section{Sequential Monte Carlo for Bayesian inversion}
Since we consider a highly nonlinear model, an iterative approach to Bayesian inversion is essential. In the framework of SMC it is performed by tempering (or annealing), when the prior measure bridged to the posterior measure not at once but through tempered measures. It should be noted that the number of tempered measures is not predefined, which could be a potential computational burden. In order to avoid filter degeneracy both resampling and mutation (or jittering) has to be performed. 
In the "classical" approach we perform monomial resampling, which we propose to replace by resampling based on optimal transport. 

\subsection{Adaptive SMC}\label{ad_SMC}
The SMC approach to Bayesian inversion involves bridging the prior $\mu_{0}$ and the posterior $\mu$ via a sequence of intermediate artificial measures $\{ \mu_{n}\}_{n=0}^{N}$, with $\mu_{N}=\mu$, defined by
\begin{eqnarray}\label{eq16}
\frac{\rmd \mu_{n}}{\rmd\mu_{0}}(u) \propto l_{n}(u,y)\equiv l(u,y)^{\phi_{n}}
\end{eqnarray}
where $\{\phi_{n}\}_{n=0}^{N}$ is a set of tempering parameters that satisfy $0=\phi_{0}<\phi_{1}<\cdots< \phi_{N}=1$. 
Expression (\ref{eq16}) formally implies
\begin{eqnarray}\label{eq17}
 \frac{\rmd \mu_{n}}{\rmd \mu_{n-1}}(u)=\frac{1}{Z_{n}}  l(u,y)^{(\phi_{n}-\phi_{n-1})}
\end{eqnarray}
where
\begin{eqnarray}\label{eq18}
Z_{n}\equiv \int_{X}  l(u,y)^{(\phi_{n}-\phi_{n-1})} \mu_{n-1}(\rmd u)
\end{eqnarray}
Let us then assume that at the iteration level $n-1$, the tempering parameter $\phi_{n-1}$ has been specified, and that a set of particles  $\{u_{n-1}^{(j)}\}_{j=1}^{J}$ provides the following approximation (with equal weights) of the intermediate measure $\mu_{n-1}$:
\begin{equation}\label{eq19}
\mu_{n-1}^{J}(u)\equiv \frac{1}{J}\sum_{j=1}^{J} \delta_{u_{n-1}^{(j)}}(u)\simeq \mu_{n-1}(u).
\end{equation}
Then from (\ref{eq18}) it follows that 
\begin{eqnarray}\label{eq20}
Z_{n}\simeq \sum_{j=1}^{J}  l(u_{n-1}^{(j)},y)^{(\phi_{n}-\phi_{n-1})} 
\end{eqnarray}
and thus, for any measureable $f$, we have that 
\begin{eqnarray}\label{eq21}
\mathbb{E}^{\mu_{n}}(f(u))\equiv \int_{X}f(u)\mu_{n}(du) =\frac{1}{Z_{n}}\int_{X}f(u) l(u,y)^{(\phi_{n}-\phi_{n-1})}\mu_{n-1}(\rmd u)\nonumber\\
 \simeq \Big[\sum_{j=1}^{J}  l(u_{n-1}^{(j)},y)^{(\phi_{n}-\phi_{n-1})} \Big]^{-1}\sum_{j=1}^{J} l(u_{n-1}^{(j)},y)^{(\phi_{n}-\phi_{n-1})} f(u_{n-1}^{(j)}),\nonumber\\
=\sum_{j=1}^{J}W_{n}^{(j)}f(u^{(j)}),\end{eqnarray}
where the importance weights for the approximation of $\mu_{n}$ are given by
\begin{eqnarray}\label{eq22}
W_{n}^{(j)} =\mathcal{W}_{n-1}^{(j)}[\phi_{n}] \equiv  \frac{l(u_{n-1}^{(j)},y)^{\phi_{n}-\phi_{n-1}}}{\sum_{s=1}^{J}l(u_{n-1}^{(s)},y)^{\phi_{n}-\phi_{n-1}}}.
\end{eqnarray}
From (\ref{eq21}) we see that the importance (normalized) weights $W_{n}^{(j)}$ assigned to each particle $u_{n-1}^{(j)}$ define the following empirical (particle) approximation of $\mu_{n}$:
\begin{eqnarray}\label{eq23}
\mu_{n}^{J}(u)\equiv \sum_{j=1}^{J}W_{n}^{(j)} \delta_{u_{n-1}^{(j)}}(u).
\end{eqnarray}

\subsubsection{Selection-Resampling Step}\label{res}

From the previous subsection it follows that adaptive SMC requires then to select the tempering parameters $\phi_{n}$ so that the two  consecutive measures $\mu_{n-1}$ and $\mu_{n}$ are sufficiently close for the IS approximating to be accurate. To this end, a common procedure  \cite{Ajay} involves imposing a threshold on the effective sample size (ESS) defined by 
\begin{eqnarray}\label{eq24}
{\rm ESS}_{n}(\phi)\equiv  \Bigg[\sum_{j=1}^{J}(\mathcal{W}_{n-1}^{(j)}[\phi])^2\Bigg]^{-1}
\end{eqnarray}
which, in turn, provides a measure of the quality of the population. In other words, $\phi_n$ is defined by the solution to 
\begin{eqnarray}\label{eq25}
{\rm ESS}_{n}(\phi)= J_{\rm thresh},
\end{eqnarray}
for a user-defined parameter $J_{\rm thresh}$ on the ESS. A bisection algorithm on the interval $(\phi_{n-1},1]$ can be used to solve (\ref{eq25}) \cite{EnsembleYO}. If ${\rm ESS}_{n}(1)>J_{\rm thresh}$, then then we can simply set $\phi_{n}=1$ as no further tempering is thus required. 

Once the tempering parameter $\phi_{n}$ has been computed via (\ref{eq25}), normalised weights (\ref{eq22}) can be computed. Since some of these can be very low, resampling with replacement according to these weights is then required to discard particles associated with those low weights. After resampling, a new set of equally-weighted particles denoted by $\hat{u}_{n}^{(j)}$ ($j=1,\dots, J$) provide a particle approximation of the measure $\mu_n$.

\subsubsection{Mutation Phase}

In order to add diversity to the resampled particles $\hat{u}_{n}^{(j)}$ computed in the selection-resampling step, a mutation step is included in most SMC methodologies. This mutation consists of sampling from a Markov kernel $\mathcal{K}_{n}$ with invariant distribution $\mu_{n}$. This can be achieved by running $N_{\mu}$ steps of an MCMC algorithm that has target distribution equal to $\mu_{n}$. An example of MCMC suitable for the parameterisation P1 of section \ref{perm} is the preconditioned Crank-Nicolson (pcn)-MCMC  \cite{CRSW13} displayed in Algorithm \ref{MCMC_al}. This algorithm samples from the target $\mu_{n}$ with reference measure $\mu_{0}=N(m,C)$; we recall these two measures are related by (\ref{eq14}). The resulting particles denoted by $\{u_{n}^{(j)}\}_{j=1}^{J}$ ($u_{n}^{(j)}\sim \mathcal{K}_{n}(\hat{u}_{n}^{(j)},\cdot)$) provide a particle approximation of $\mu_{n}$ in the form
\begin{eqnarray}\label{eq39}
\mu_{n}^{J}\equiv \frac{1}{J}\sum_{j=1}^{J} \delta_{u_{n}^{(j)}}
  \end{eqnarray}
Convergence of (\ref{eq39}) to $\mu_{n}$ in the large ensemble size limit can be found in  \cite{BJMS15}. The complete adaptive SMC sampler is displayed in Algorithm~\ref{SMC_al}.

\begin{algorithm}[H]
\caption{\small pcn-MCMC to generate samples from a $\mu_{n}$-invariant Markov kernel with $\mu_{0}=N(m,C)$}\label{MCMC_al}
\begin{algorithmic}
\Statex Select $\beta\in (0,1)$ and an integer $N_{\mu}$.
\For{ $j=1,\dots, J$}
\Statex Initialize $\nu^{(j)}(0)=  \hat{u}_{n}^{(j)}$
  \While{$\alpha\leq N_{\mu}$}
\State{(1) \textbf{pcN proposal}. Propose $u_{\rm prop}$ from
\begin{eqnarray}\label{prop_ape}
u_{\rm prop}=\sqrt{1-\beta^2}\nu^{(j)}(\alpha)+(1-\sqrt{1-\beta^2})m+\beta\xi, \qquad \textrm{with}~~ \xi\sim N(0,C) \nonumber
\end{eqnarray}}
\State{(2) Set $\nu^{(j)}(\alpha+1)=u_{\rm prop}$ with probability $a(\nu^{(j)}(\alpha),u)$ and  $\nu^{(j)}(\alpha+1)=\nu^{(j)}(\alpha)$ with probability $1-a(\nu^{(j)}(\alpha),u)$, where
\begin{eqnarray}\label{eq:3.6}
a(u,v)=\min\Big\{1,  \frac{ l(u,y )^{\phi_{n}}}{l(v,y)^{\phi_{n}}  }\Big\} \nonumber, \ \mbox{with}\ l\ \mbox{defined in (\ref{eq12})}
\end{eqnarray}}
\State{(3) $\alpha \gets \alpha+1$}
  \EndWhile
\EndFor
\end{algorithmic}
\end{algorithm}

\begin{algorithm}[H]
\caption{SMC algorithm for High-Dimensional Inverse Problems}\label{SMC_al}
\begin{algorithmic}
\Statex Let $\{u_{0}^{(j)}\}_{j=1}^{J}\sim \mu_{0}$ be the initial ensemble of $J$ particles.
\Statex Define the tunable parameters $J_{\rm thresh}$ and $N_{\mu}$.
\State Set $n=0$ and $\phi_{0}=0$
\While{ $\phi_{n}<1$}
\State $n\to n+1$
\State  \textbf{Compute the likelihood} (\ref{eq12}) $l(u_{n-1}^{(j)},y)$ (for $j=1,\dots, J$)
\State   \textbf{Compute the tempering parameter $\phi_{n}$}:
\If{ $\min_{\phi\in (\phi_{n-1},1)}\textrm{ESS}_{n}(\phi)>J_{\rm thresh}$}
\State set $\phi_{n}=1$.
\Else
\State compute $\phi_{n}$ such that $\textrm{ESS}_{n}(\phi)\approx J_{\rm thresh}$
\State using a bisection algorithm on $(\phi_{n-1},1]$.
\EndIf
\State \textbf{Computing weights} from expression (\ref{eq22}) $W_{n}^{(j)}\equiv \mathcal{W}_{n-1}^{(j)}[\phi_{n}]$
\State \textbf{Resample}. Let $(p^{(1)},\dots,p^{(J)})\in \mathcal{R}(W_{n}^{(1)},\dots,W_{n}^{(J)})$, 
where $\mathcal{R}$ denotes multinomial resampling with replacement.
\State  Set $\hat{u}_{n}^{(j)}\equiv u_{n-1}^{(p^{(j)})}$ and $W_{n}^{(j)}=\frac{1}{J}$
\State\textbf{Mutation}. Sample $u_{n}^{(j)}\sim \mathcal{K}_{n}(\hat{u}_{n}^{(j)},\cdot )$ via Algorithm \ref{MCMC_al}.
\EndWhile
\State Approximate $\mu_{n}$ by $\mu_{n}^{J}\equiv \frac{1}{J}\sum_{j=1}^{J} \delta_{u_{n,r}^{(j)}}$
\end{algorithmic}
\end{algorithm}

\subsection{Optimal Transport within SMC}
In this section we assume that $X=\mathbb{R}^{K}$. We denote $U_{n-1}$ a discrete random variable with realisations $\{u_{n-1}^{(j)}\}_{j=1}^{J}$ and probabilities $\{W_{n}^{(j)}\}_{j=1}^{J}$. We denote $U_{n}$ the random variable with samples $\{\hat{u}_{n-1}^{(j)}\}_{j=1}^{J}$ with equal weights. 
The aim is to replace the resampling step in the method above with 
resampling that maximizes the covariance between $U_{n-1}$ and $U_{n}$. 
Such a resampling is performed by finding a coupling between the posterior defined by the weights  $\{W_{n}^{(j)}\}_{j=1}^{J}$ and the uniform probability density such that it maximizes the covariance between $U_{n-1}$ and $U_{n}$.

\mi{
Let us assume that the two consecutive measures $\mu_{n-1}$ and $\mu_n$ are defined on a measurable space $(\Omega,\mathcal{F})$ such that $\mu_{n-1}$ is the law of $U_{n-1}: \Omega \rightarrow \mathcal{U}_{n-1}$ and $\mu_n$ is the law of $U_n:  \Omega \rightarrow \mathcal{U}_{n}$. 
Here, the couple $(U_{n-1},U_n)$ is called the coupling of $(\mu_{n-1},\mu_n)$, i.e. the coupling of the posterior defined by the weights  $\{W_{n}^{(j)}\}_{j=1}^{J}$ and  the uniform probability density. A coupling is called deterministic if there exists a measurable function $\Psi: \mathcal{U}_{n-1} \rightarrow \mathcal{U}_n$ such that $U_n = \Psi (U_{n-1})$ and $\Psi$ is called transport map. 
Unlike couplings, deterministic couplings do not always exist.  On the other hand there may be an infinitely many deterministic couplings.
An example of deterministic coupling is an optimal coupling.
Optimal coupling is a solution of 
the Monge-Kantorovitch miminization problem 
\[
 \inf\int_{\mathcal{U}_{n-1}\times\mathcal{U}_n} c(u_{n-1},\hat{u}_{n-1}) \rmd\ell(u_{n-1},\hat{u}_{n-1}),
\]
where minimum runs over all joint probability measures $\ell$ on $\mathcal{U}_{n-1}\times\mathcal{U}_n$ with marginals $\mu_{n-1}$ and $\mu_n$,
and $c(u_{n-1},\hat{u}_{n-1})$ is a cost function on $\mathcal{U}_{n-1}\times\mathcal{U}_n$.
The joint measures achieving the infinum are called optimal transference plans. The optimal coupling is unique if
the measure $\mu_{n-1}$ possess some regularity properties and the cost function $c(u_{n-1},\hat{u}_{n-1})$ is convex~\cite{Villani}.
It appeared that such a coupling simultaneously 
minimizes the expectation between $||u_{n-1} - \hat{u}_{n}||^2$ and is defined as the solution of the Monge-Kantorovitch problem with cost function $ c(u_{n-1},\hat{u}_{n})= ||u_{n-1}-\hat{u}_{n}||^2$. 
}
Thus the above described coupling is a $J\times J$ matrix $T^{*}$ with non-negative entries $T_{ij}^{*}$ that satisfy
\begin{equation}\label{eq:OTc1}
 \sum_{i=1}^J T_{ij}^{*} = \frac{1}{J}, \qquad   \sum_{j=1}^J T_{ij}^{*} = W_{i},
\end{equation}
and minimizes 
\begin{equation}\label{eq:OTc2}
 \sum_{i,j=1}^{J}T_{ij}||u_{n-1}^{(i)}-\hat{u}_{n}^{(j)}||^2\equiv \sum_{i,j=1}^{J}T_{ij}||u_{n-1}^{(i)}-u_{n-1}^{(j)}||^2
\end{equation}
for $T_{ij}^{*} $. This is a {\it linear} transport problem of finding $J^2$ unknowns.
Then the linear transformation gives new samples according to 
\begin{equation}\label{eq:OT}
\hat{u}_{n}^{(j)}:=\sum_{i=1}^{J} P_{ij}u_{n-1}^{(j)} \qquad  \mbox{for} \  j=1,\dots,J,
\end{equation}
where $P_{ij}=JT_{ij}^{*}$. 

The deterministic optimal transformation~(\ref{eq:OT}) converges weakly to 
the solution of the underlying continuous Monge-Kantorovitch problem as $J\to\infty$~\cite{Re13}.
ETPF is first order consistent, since
$$ \overline{\hat{u}}_{n}=\frac{1}{J}\sum_{j=1}^{J}\hat{u}_{n}^{(j)}=\frac{1}{J}\sum_{j=1}^{J}\sum_{i=1}^{J} P_{ij}u_{n-1}^{(j)}=\sum_{j=1}^{J}\sum_{i=1}^{J} T_{ij}^{*}u_{n-1}^{(j)}=\sum_{j=1}^{J}W_{n}^{(j)}u_{n-1}^{(j)}.$$
There also exists a second-order accurate ETPF~\cite{Wietal17}, which however does not satisfy $T_{ij}^{*}\geq 0$.
The main difference between resampling based on optimal transport  and monomial resampling  
is that the former one is optimal in the sense of the Monge-Kantorovitch problem,
while the latter one is non-optimal in that sense.

The computational complexity of finding the minimizer of~(\ref{eq:OT}) is in general $\Or(J^3\ln J)$, 
which has been reduced to $\Or(J^2\ln J)$ in~\cite{PeWe09}.
\mi{The wall clock time at $J=100$ is 0.3 seconds  for SMC with optimal resampling,
while 0.03 seconds for both SMC with monomial resampling and EKI. }
It can be further improved by employing fast iterative methods for finding approximate minimizers using the Sinkhorn distance \cite{Cuturi13}, which was implemented in~\cite{Wietal17} for the second-order accurate ETPF.
The algorithm of Earth's moving distances of~\cite{PeWe09} is available as both {\it MATLAB} and {\it Python} codes
and is used here. The complete adaptive optimal transport based SMC sampler is displayed in Algorithm~\ref{SMC_opt}.

\begin{algorithm}[H]
\caption{Optimal transport based SMC algorithm for High-Dimensional Inverse Problems}\label{SMC_opt}
\begin{algorithmic}
\Statex Let $\{u_{0}^{(j)}\}_{j=1}^{J}\sim \mu_{0}$ be the initial ensemble of $J$ particles.
\Statex Define the tunable parameters $J_{\rm thresh}$ and $N_{\mu}$.
\State Set $n=0$ and $\phi_{0}=0$
\While{ $\phi_{n}<1$}
\State $n\to n+1$
\State  \textbf{Compute the likelihood} (\ref{eq12}) $l(u_{n-1}^{(j)},y)$ (for $j=1,\dots, J$)
\State   \textbf{Compute the tempering parameter $\phi_{n}$}:
\If{ $\min_{\phi\in (\phi_{n-1},1)}\textrm{ESS}_{n}(\phi)>J_{\rm thresh}$}
\State set $\phi_{n}=1$.
\Else
\State compute $\phi_{n}$ such that $\textrm{ESS}_{n}(\phi)\approx J_{\rm thresh}$
\State using a bisection algorithm on $(\phi_{n-1},1]$.
\EndIf

\State \textbf{Computing weights} from expression (\ref{eq22}) $W_{n}^{(j)}\equiv \mathcal{W}_{n-1}^{(j)}[\phi_{n}]$
\State \textbf{Resample based on optimal transport}. Compute $\mathcal{D}_{ij} = ||u_{n-1}^{(i)}-u_{n-1}^{(j)}||^2$  (for $i,j=1,\dots, J$). Supply $\{\mathcal{D}_{ij}\}_{i,j=1}^{J}$ and $\{W_{n}^{(j)}\}_{j=1}^{J}$ to the Earth's moving distances algorithm of Pele \& Werman. The output is the coupling $\{T_{ij}^{*}\}_{i,j=1}^{J}$.
\State  Compute new samples  $\hat{u}_{n}^{(j)}$ (\ref{eq:OT})  and set $W_{n}^{(j)}=\frac{1}{J}$.

\State\textbf{Mutation}. Sample $u_{n}^{(j)}\sim \mathcal{K}_{n}(\hat{u}_{n}^{(j)},\cdot )$ via Algorithm \ref{MCMC_al}.
\EndWhile
\State Approximate $\mu_{n}$ by $\mu_{n}^{J}\equiv \frac{1}{J}\sum_{j=1}^{J} \delta_{u_{n,r}^{(j)}}$

\end{algorithmic}
\end{algorithm}

\subsection{Gaussian Approximation of SMC via ensemble Kalman inversion}

A natural approximation that arises from the adaptive SMC framework described in subsection \ref{ad_SMC} involves ensemble Kalman inversion (EKI) \cite{RTM}. More specifically, let us assume that at the $n-1$ iteration level, we approximate $\mu_{n-1}$ with a Gaussian $\hat{\mu}_{n-1}=N(m_{n-1},C_{n-1})$ where the mean $m_{n-1}$ and covariance $C_{n-1}$ are the empirial mean and covariance of the particles (assumed with equal weights) at the current iteration level. That is, 
\begin{eqnarray}\label{EKI1}
\fl m_{n-1}\equiv \frac{1}{J}\sum_{j=1}^{J}u_{n-1}^{(j)},\qquad C_{n-1}\equiv  \frac{1}{J-1}\sum_{j=1}^{J}(u_{n-1}^{(j)}-m_{n-1})\otimes (u_{n-1}^{(j)}-m_{n-1})
\end{eqnarray}
If we now linearise the forward map around $m_{n-1}$ and replace Frechet derivatives of the forward map with covariances/crosscovariances as in \cite{EnsembleYO}, it can be shown that the application to Bayes rule yields an approximate posterior $\hat{\mu}_{n}=N(m_{n},C_{n})$ with mean and covariance given by
\begin{eqnarray}\label{EKI2}
m_{n}=m_{n-1}+  C_{n-1}^{u\cG}(C_{n-1}^{\cG \cG}+\alpha_{n}\Gamma)^{-1}(y-\overline{\cG}_{n-1}),\\
C_{n}=C_{n-1}-  C_{n-1}^{u\cG}(C_{n-1}^{\cG \cG}+\alpha_{n}\Gamma)^{-1} C_{n-1}^{\cG u},
\end{eqnarray}
where 
\begin{eqnarray}\label{EKI3}
\fl \overline{\cG}_{n-1}\equiv \frac{1}{J}\sum_{j=1}^{J}\cG(u_{n-1}^{(j)}),\qquad C_{n-1}^{u\cG}\equiv  \frac{1}{J-1}\sum_{j=1}^{J}(u_{n-1}^{(j)}-m_{n-1})\otimes (\cG(u_{n-1}^{(j)})-\overline{\cG}_{n-1}),\\
 C_{n-1}^{\cG\cG}\equiv  \frac{1}{J-1}\sum_{j=1}^{J}(\cG(u_{n-1}^{(j)})-\overline{\cG}_{n-1})\otimes (\cG(u_{n-1}^{(j)})-\overline{\cG}_{n-1}),
\end{eqnarray}
and where 
\begin{eqnarray}\label{EKI4}
\alpha_{n}=\frac{1}{\phi_{n}-\phi_{n-1}}.
\end{eqnarray}
Since we are interested in a particle approximation of $\hat{\mu}_{n}=N(m_{n},C_{n})$, we can use the following expression 
\begin{eqnarray}\label{EKI5}
\hat{u}_{n}^{(j)} =u_{n-1}^{(j)}+C_{n-1}^{u\cG}(C_{n-1}^{\cG \cG}+\alpha_{n}\Gamma)^{-1}(y_{n}^{(j)}-\cG_{n}(u_{n-1}^{(j)})),
\end{eqnarray}
where 
\begin{eqnarray}\label{EKI6}
y_{n}^{(j)}\equiv y+\eta_{n}^{(j)},\qquad \eta_{n}^{(j)}\sim N(0,\alpha_{n}\Gamma_{n}).
\end{eqnarray}
Standard Kalman filter arguments \cite{0266-5611-5-4-011} can be used to show that the particle approximation provided by (\ref{EKI5})-(\ref{EKI6}) converges to $\hat{\mu}_{n}$ as $J\to \infty$. We note in passing that, within the adaptive SMC framework used here, the regularisation/inflation parameter $\alpha_{n}$ in formulas (\ref{EKI4}) is computed based on the ESS criteria discussed in subsection \ref{res}.

It is important to emphasize that, in general, the approximate Gaussian measure $\hat{\mu}_{n}$ coincides with $\mu_{n}$ only when the forward map is linear and the prior $\mu_{0}$ is Gaussian. The approximation provided by EKI will deteriorate when we depart from Gaussian-linear assumptions. Therefore, we propose to conduct MCMC mutations to each of the particles in (\ref{EKI5}) with the aim of improving the approximation of each posterior measure $\mu_{n}$. The complete EKI-based algorithm is displayed in Algorithm \ref{EKI_Al}. \mi{We recognise that this is only an ad-hoc approach for which exact sampling of the posterior (as $J\to \infty$) is not ensured. A more rigorous (i.e. fully-Bayesian approach) that we leave for future work is to use EKI in the proposal design for the importance sampling step within SMC; this is done for data assimilation settings in \cite{Chetal16}. }

\begin{algorithm}[H]
\caption{EKI approximation to SMC }\label{EKI_Al}
\begin{algorithmic}
\Statex Let $\{u_{0}^{(j)}\}_{j=1}^{J}\sim \mu_{0}$ be the initial ensemble of $J$ particles.
\Statex Define the tunable parameters $J_{\rm thresh}$ and $N_{\mu}$.
\State Set $n=0$ and $\phi_{0}=0$
\While{ $\phi_{n}<1$}
\State $n\to n+1$
\State  \textbf{Compute the likelihood} (\ref{eq12}) $l(u_{n-1}^{(j)},y)$ (for $j=1,\dots, J$)
\State   \textbf{Compute the tempering parameter $\phi_{n}$}:
\If{ $\min_{\phi\in (\phi_{n-1},1)}\textrm{ESS}_{n}(\phi)>J_{\rm thresh}$}
\State set $\phi_{n}=1$.
\Else
\State compute $\phi_{n}$ such that $\textrm{ESS}_{n}(\phi)\approx J_{\rm thresh}$
\State using a bisection algorithm on $(\phi_{n-1},1]$.
\EndIf

\State Generate particles $\{\hat{u}_{n}^{(j)}\}_{j=1}^{J}$ according to (\ref{EKI5}). 

\State\textbf{Mutation}. Sample $u_{n}^{(j)}\sim \mathcal{K}_{n}(\hat{u}_{n}^{(j)},\cdot )$ via Algorithm \ref{MCMC_al}.
\EndWhile
\State Approximate $\mu_{n}$ by $\mu_{n}^{J}\equiv \frac{1}{J}\sum_{j=1}^{J} \delta_{u_{n,r}^{(j)}}$

\end{algorithmic}
\end{algorithm}

\section{Numerical experiments}
In this section we perform numerical experiments to infer P1 and P2 parameters.
We compare optimal transport based SMC to both monomial based SMC and EKI, 
which we denote optimal, monomial, and Kalman, respectively. 
We analyze methods performance with respect to a pcn-MCMC solution, which we denote as reference.
We combine 50 independent chains each of the length $10^6$ and $10^5$ burn-in period and thinning $10^3$.

Observations of pressure were obtained from the true permeability 
with observation noise from normal distribution with zero mean and standard deviation of 2\% of $L^2$-norm of the true pressure.
We should note that both the true random variable and an initial ensemble of parameterized permeability are drawn from the same prior distribution as the prior includes knowledge about geological properties. 
However, the true solution is computed on a fine grid and an initial guess on a coarse grid, which is half the resolution of the fine grid.
The uncertain parameter for P1 inference has the dimension of the coarse grid, i.e. $4900=70^2$.
The uncertain parameter for P2 inference has the dimension 
of the coarse grid twice, due to permeability defined inside and outside channel but on the whole grid,
plus the dimension of the geometrical parameters, i.e. $5005 = 50^2 + 50^2 +5$.

For log-permeability parameters, the prior is normal distribution with mean 5 for P1,
and for P2 with mean 15 outside channel and 100 inside channel.
For geometrical parameters, the prior is uniform: $d_1\sim U[0.05\times6,\ 0.35\times6]$,
$d_2\sim U[\pi/2,\ 6\pi]$,
$d_3\sim U[-\pi/2,\ \pi/2]$,
$d_4\sim U[0,\ 6]$,
$d_5\sim U[0.02\times6,\ 0.7\times6]$.
For tempering we choose the effective ensemble size threshold $J_{\rm thresh} = J/3$ 
and for mutations the length of Markov chain $N_{\mu}=10$ to save computational costs.
For P2, we use Metropolis-within-Gibbs methodology of~\cite{ILS14} to separate geometrical parameters and 
log-permeability parameters within the mutation step, since it allows to better exploit the structure of the prior.
The proposal design for the geometric parameters within the Metropolis-within-Gibs consist of local moves within the intervals of the prior with a step size that we tune to achieve acceptance rates between 20\% and 30\%. Geometrical
 parameters that fall outside those intervals are projected back via a projection that preserves reversibility of the proposal with respect to the prior~\cite{ILS14}.
We perform numerical experiments with different ensemble sizes of 100, 500, and 1000.
We perform 10 simulations with different realizations of the initial ensemble to check the robustness of results.

For log-permeability, we compute $L^2$ norm of the error in the mean with respect to the reference  
\begin{equation*}\label{eq:MSE}
{\rm Error} =||\overbar{u}-\overbar{u}^{\rm ref}||, \quad\mbox{where}\quad \overbar{u}=\frac{1}{J}\sum_{j=1}^J u^{(j)}.
\end{equation*}
\mi{We investigate the performance of the proposed approach to approximate the marginal posterior, $p(d_i)$, of each geometric parameter $d_i$ ($i=1,\dots, 5$) defined in parameterisation P2. To this end, we compute Kullback-Leibler divergence with respect to the reference/true posterior marginal (denoted by $p^{\rm ref}(d_i)$) computed via MCMC:
\begin{equation}\label{eq:KL}
{\rm D}_{\rm KL} (p^{\rm ref} \parallel p) = \sum_{j=1}^{J_{\rm b}} p^{\rm ref}(d^j_i) \log \frac{p^{\rm ref}(d^j_i)}{p(d^j_i)}, 
\end{equation}
where $J_{\rm b} = J/10$ is chosen number of bins and $p(d_i^j)$ is approximated by the weights.}
\mi{The results (median, 25 and 75 percentiles) that we report below for both the error in the mean and the KL divergence are computed over 10 experiments corresponding to independent choices of the prior ensemble.}

\subsection{Numerical inference for P1 }\label{sec:P1}
\begin{figure}[!ht]
  \centering
  \includegraphics[width=0.5\textwidth]{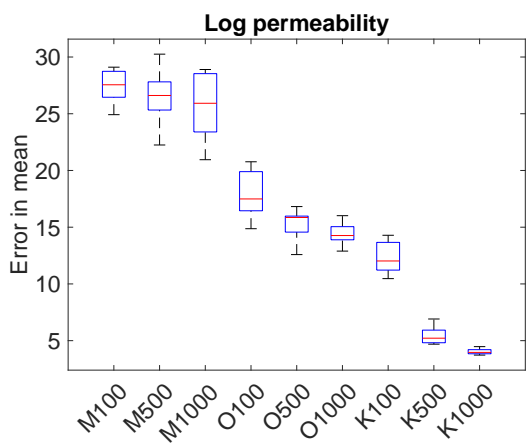}
  \caption{Box plot of the error in the mean log-permeability for P1 inference.
  Central mark is the median, edges of the box are the
    25th and 75th percentiles, whiskers extend to the most extreme
    datapoints over 10 independent simulations.
  On x-axis numbers stand for ensemble sizes, M stands for monomial based SMC, O for optimal transport based SMC, and K for EKI.  \label{fig:GausErr}}
  \centering
\end{figure}

\begin{figure}[!ht]
  \centering
  \includegraphics[width=0.9\textwidth]{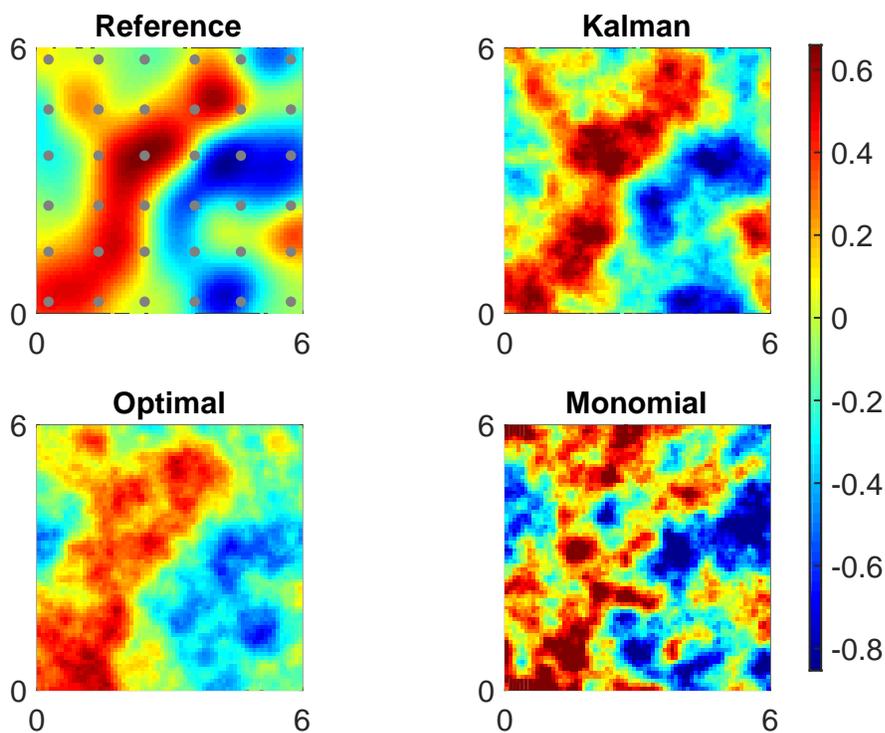}
  \caption{Mean log-permeability for P1 inference for the lowest error at ensemble size 100. Observation locations are shown in circles.
\label{fig:GausMF}}
  \centering
\end{figure}

For P1, we perform a numerical experiment using 36 uniformly distributed observations.
In Figure~\ref{fig:GausErr}, we plot error in the mean log-permeability with respect to reference.
We observe that while optimal transport based SMC outperforms monomial based SMC for all
ensemble sizes, EKI outperforms both SMC methods. This is due to the nature of P1 parametrization
and only two degrees of freedom (mean and variance) of EKI.
In Figure~\ref{fig:GausMF}, we plot mean log-permeability for a simulation with smallest error at ensemble size 100 and 
reference mean log-permeability. We see that monomial based SMC gives a less smooth estimation compared to optimal transport based SMC, EKI, and reference, which leads to larger error.

For ensemble sizes considered here, the number of tempering steps on average is 15 for optimal transport based SMC, and 17 for both monomial based SMC and EKI. Thus in terms of computational cost optimal transport based SMC is equivalent to monomial based SMC, since computational complexity of the forward model is higher than $O(J\ln J)$.

\subsection{Numerical inference for P2}\label{sec:P2}
\begin{figure}[!ht]
  \centering
  \includegraphics[width=0.9\textwidth]{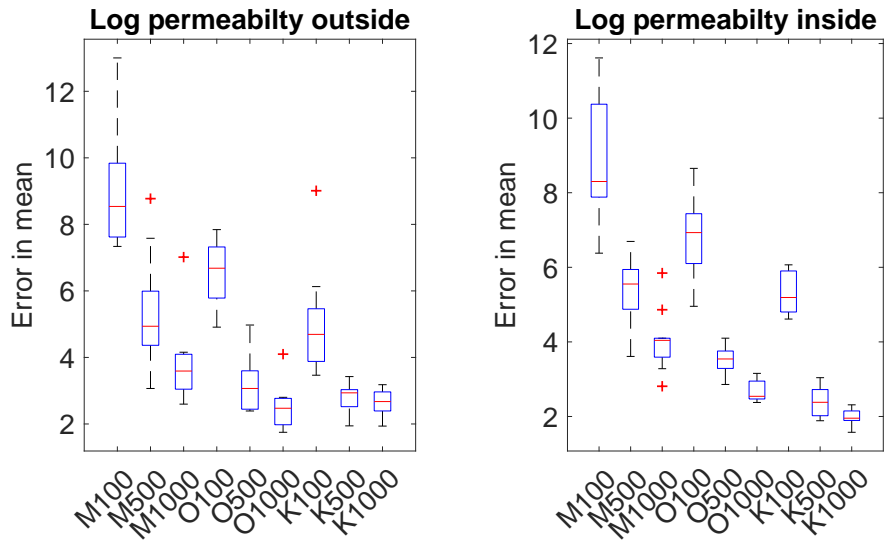}
  \caption{Box plot of the error in the mean log-permeability for P2 inference. 
Central mark is the median, edges of the box are the
    25th and 75th percentiles, whiskers extend to the most extreme
    datapoints, and crosses are outliers over 10 independent simulations.
  On the left: outside channel, on the right: inside  channel.
  On x-axis numbers stand for ensemble sizes, M stands for monomial based SMC, O for optimal transport based SMC, and K for EKI.  \label{fig:ChanErr}}
  \centering
\end{figure}

\begin{figure}[!ht]
  \centering
  \includegraphics[width=0.9\textwidth]{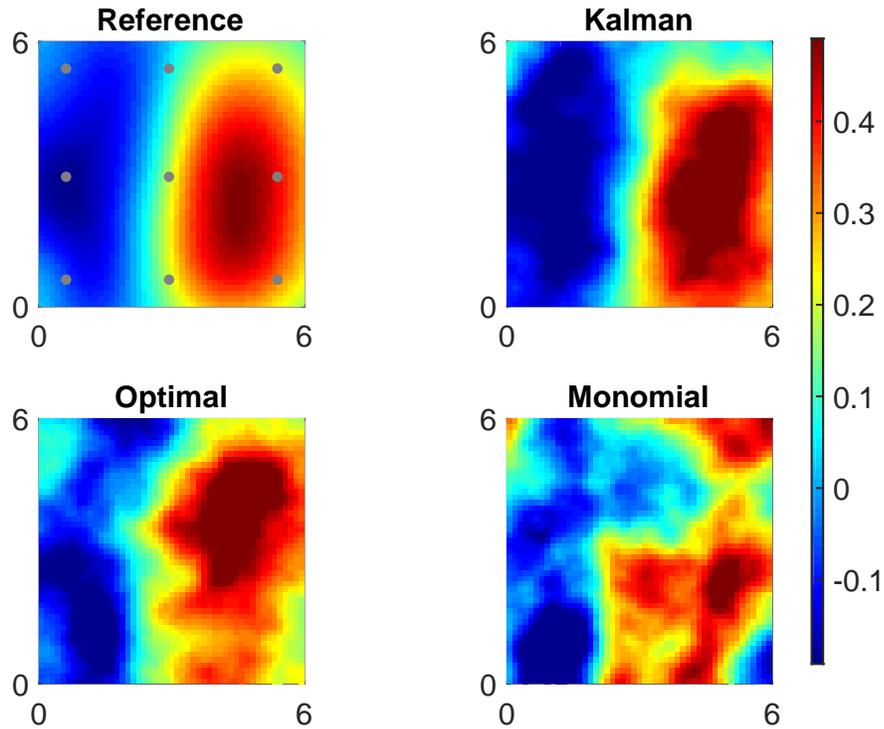}
  \caption{Mean log-permeability outside channel for P2 inference for the lowest error at ensemble size 100. Observation locations are shown in circles.
\label{fig:ChanOMF}}
 \centering
\end{figure}

\begin{figure}[!ht]
  \centering
  \includegraphics[width=0.9\textwidth]{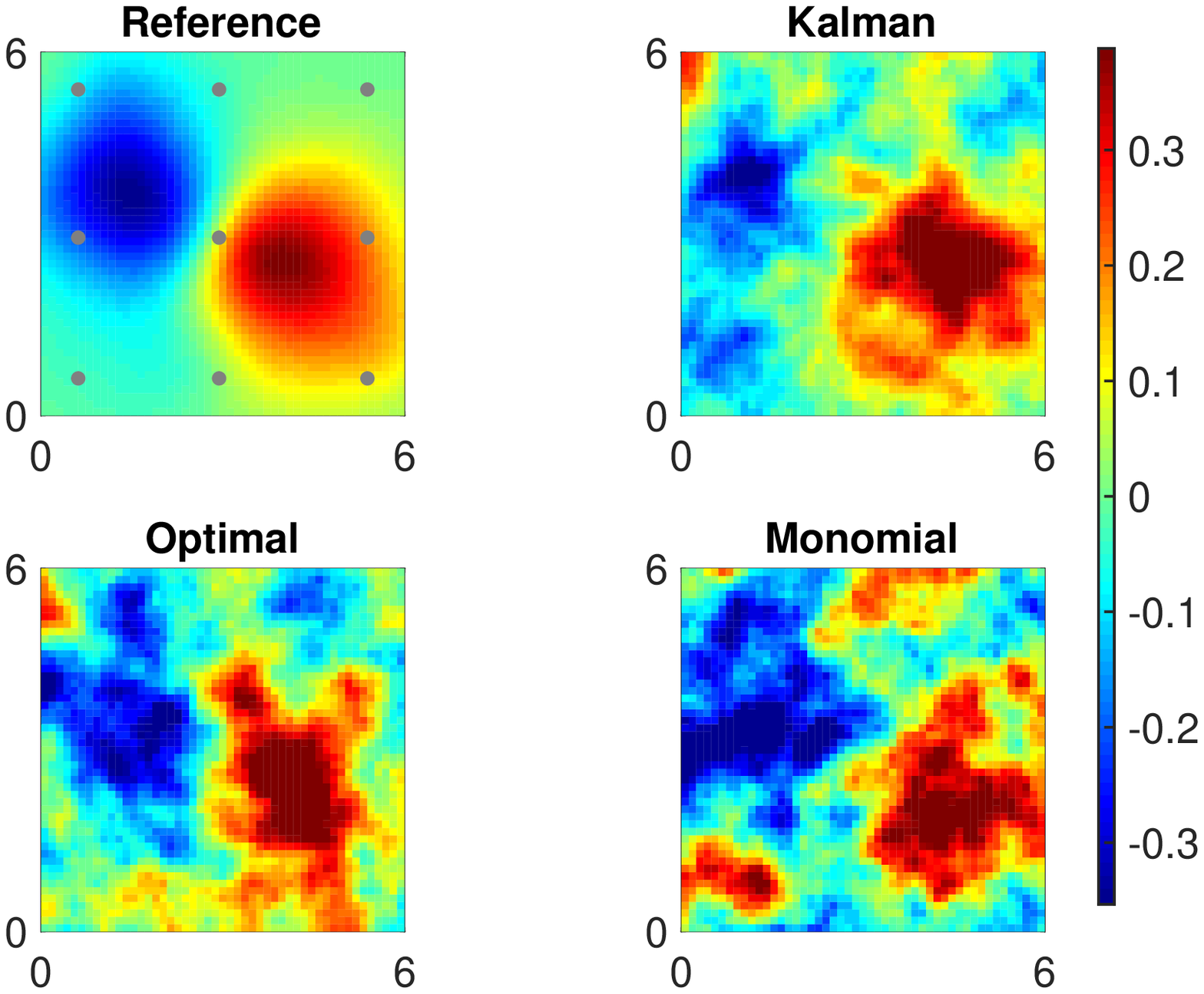}
  \caption{Mean log-permeability inside channel for P2 inference for the lowest error at ensemble size 100. Observation locations are shown in circles.
\label{fig:ChanIMF}}
  \centering
\end{figure}

For P2, we perform a numerical experiment using 9 uniformly distributed observations. For ensemble size considered here, the number of tempering steps on average is 8 for EKI, and 7 for both 
optimal transport based SMC and monomial based SMC.
In Figure~\ref{fig:ChanErr}, we plot error in the mean log-permeability with respect to reference
for permeability outside channel on the left and for permeability inside channel on the right.
We observe that while optimal transport based SMC still outperforms monomial based SMC for all
ensemble sizes, it is now comparable to EKI. This is due to a small number of observations. 
In Figures~\ref{fig:ChanOMF}--\ref{fig:ChanIMF}, we plot mean log-permeability for a simulation with smallest error at ensemble size 100 and reference mean log-permeability for permeability outside channel and for permeability inside channel, respectively. We see that monomial based SMC gives a less smooth estimation compared to optimal transport based SMC, EKI, and reference, which leads to larger error.

\begin{figure}[ht]
  \centering
  \includegraphics[width=0.9\textwidth]{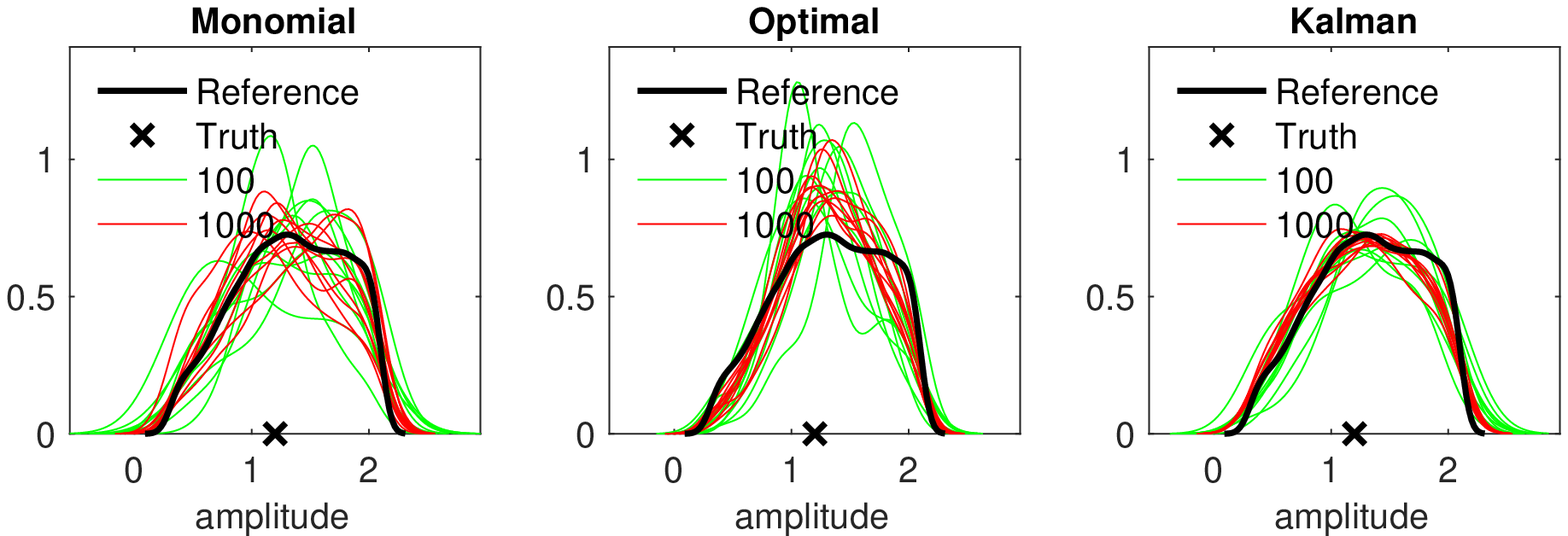}
  \includegraphics[width=0.9\textwidth]{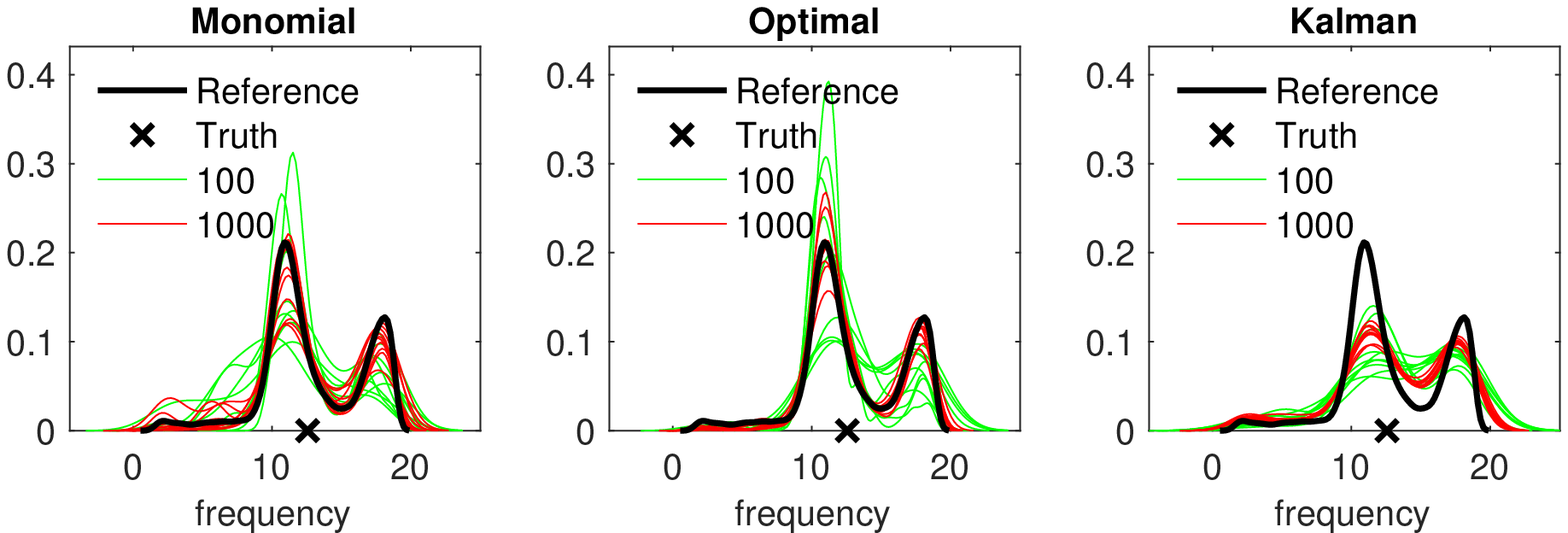}
  \includegraphics[width=0.9\textwidth]{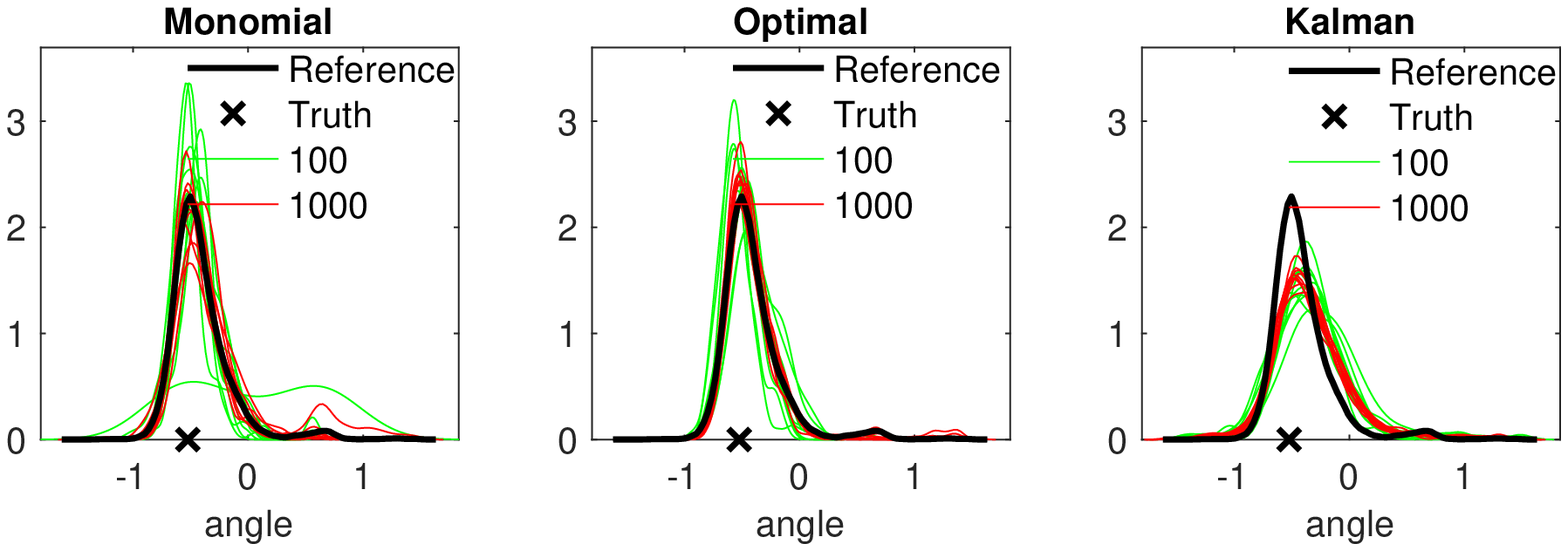}
  \includegraphics[width=0.9\textwidth]{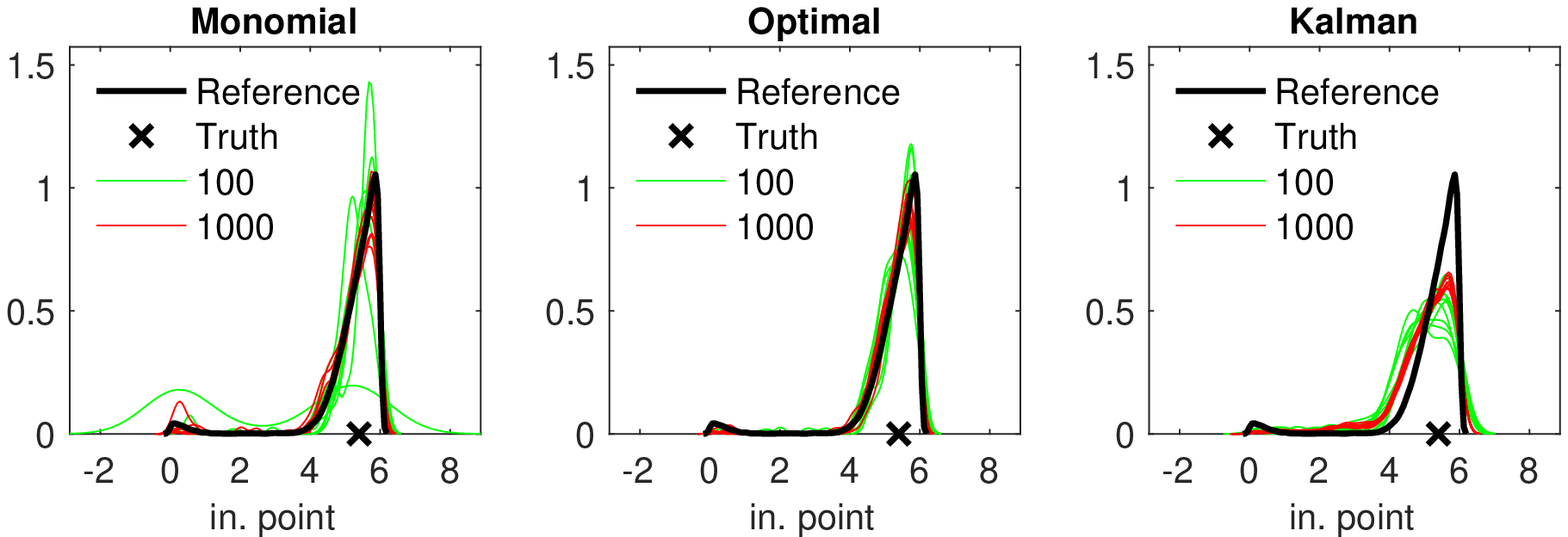}
  \includegraphics[width=0.9\textwidth]{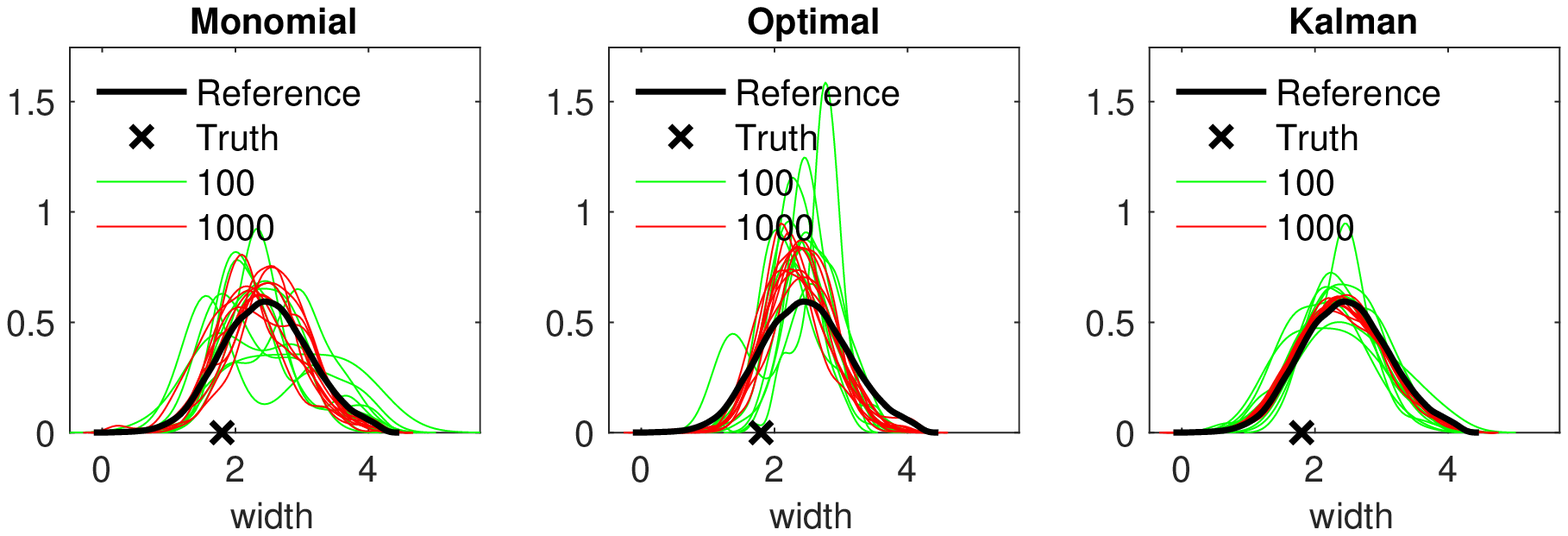}
  \caption{Posterior of geometrical parameters for P2 inference. In black is reference, in green 10 simulations of ensemble size 100, in red 10 simulations of ensemble size 1000. The true parameters are shown as black cross.
\label{fig:ChanPDF}}
  \centering
\end{figure}

\begin{figure}[ht]
  \centering
  \includegraphics[width=0.5\textwidth]{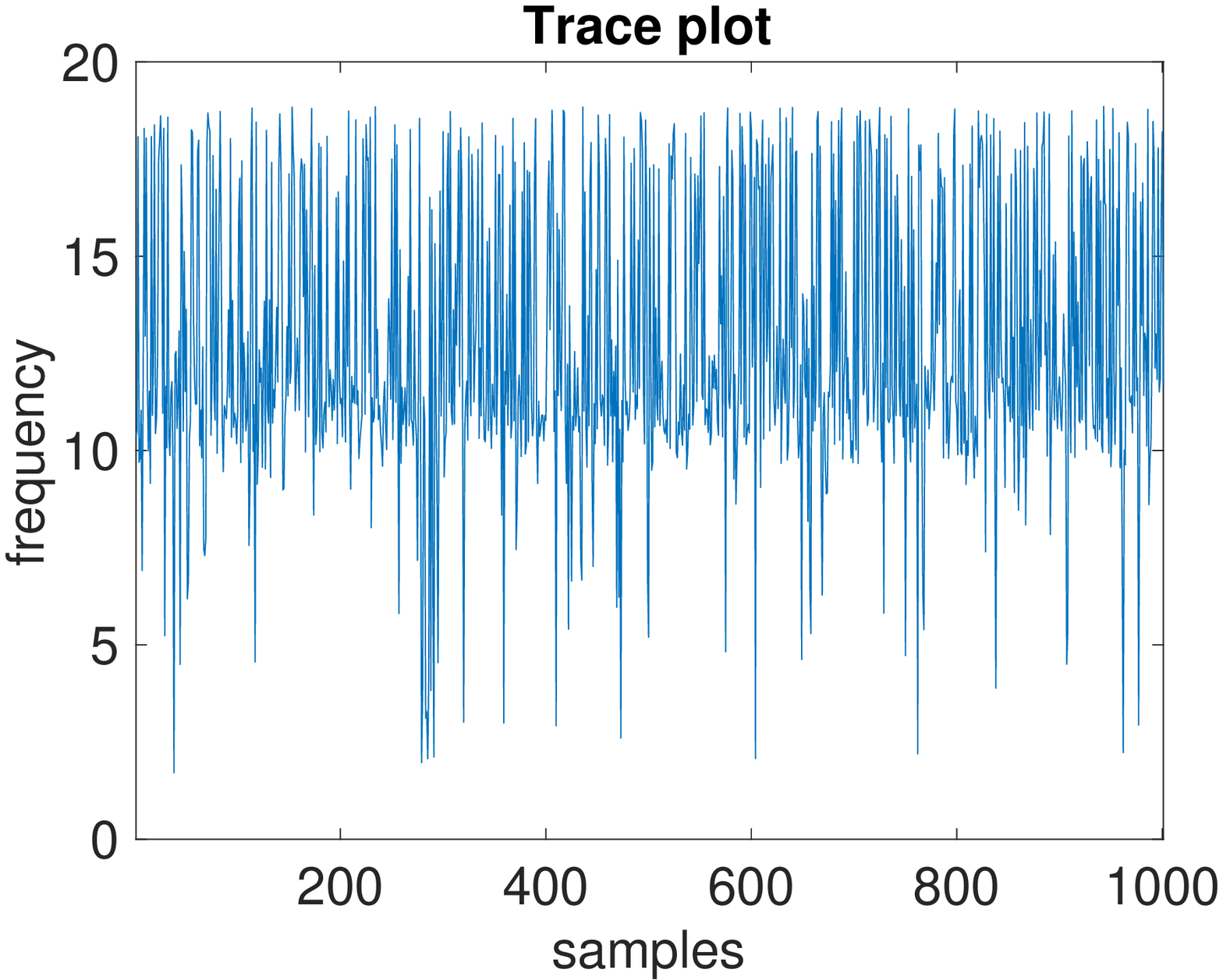}
  \caption{Trace plot of frequency from a pcn-MCMC chain.
\label{fig:ChanTP}}
  \centering
\end{figure}

In Figure~\ref{fig:ChanPDF}, we show posterior estimations of geometrical parameters. We see that all the parameters except amplitude and width exhibit strongly non-Gaussian behaviour. 
In Figure~\ref{fig:ChanTP}, we show a trace plot of frequency from a chain of the reference to check whether two modes are being sampled within each chain. We observe that the chain is properly mixed.

\begin{figure}[!ht]
  \centering
  \includegraphics[width=0.9\textwidth]{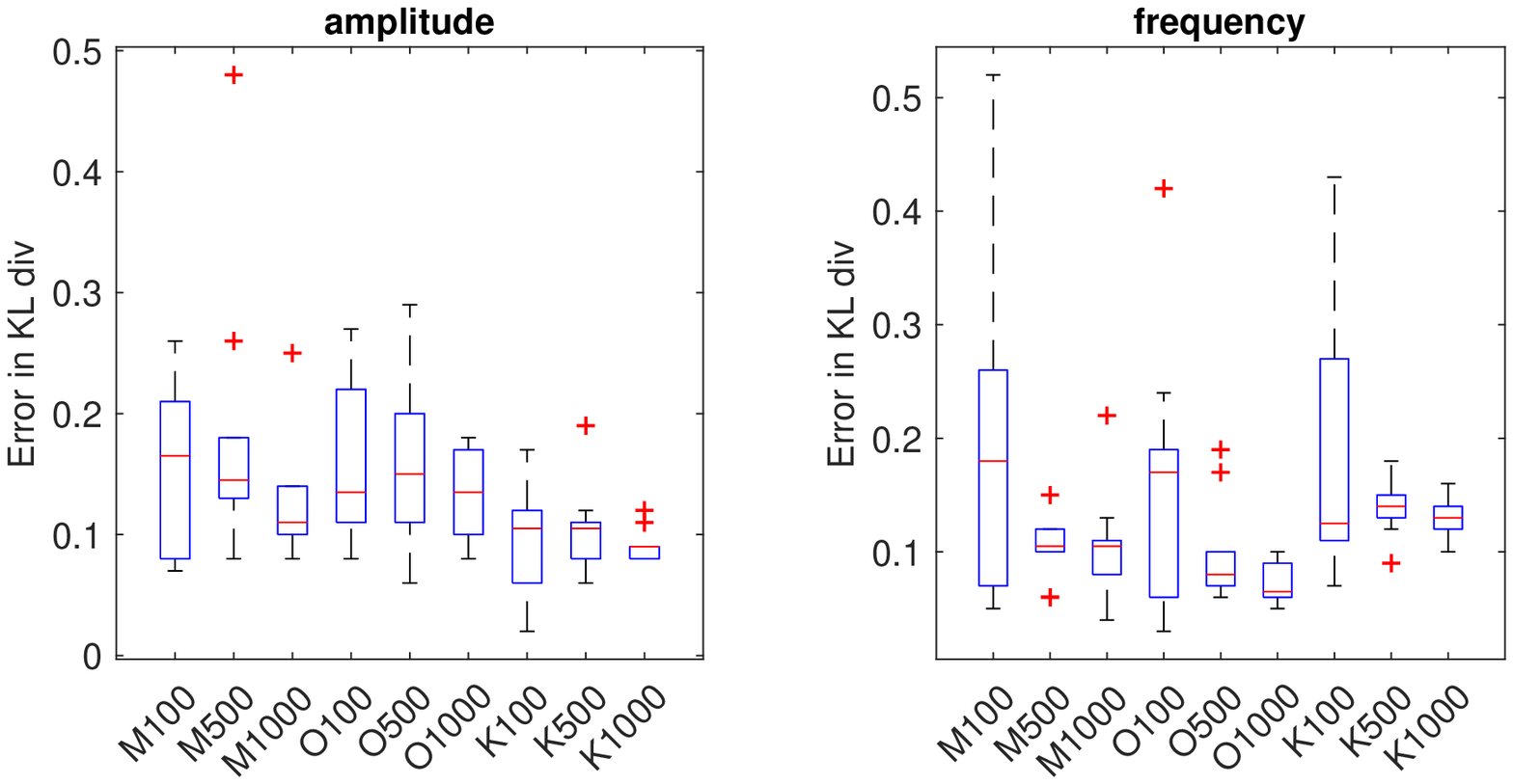}
  \includegraphics[width=0.9\textwidth]{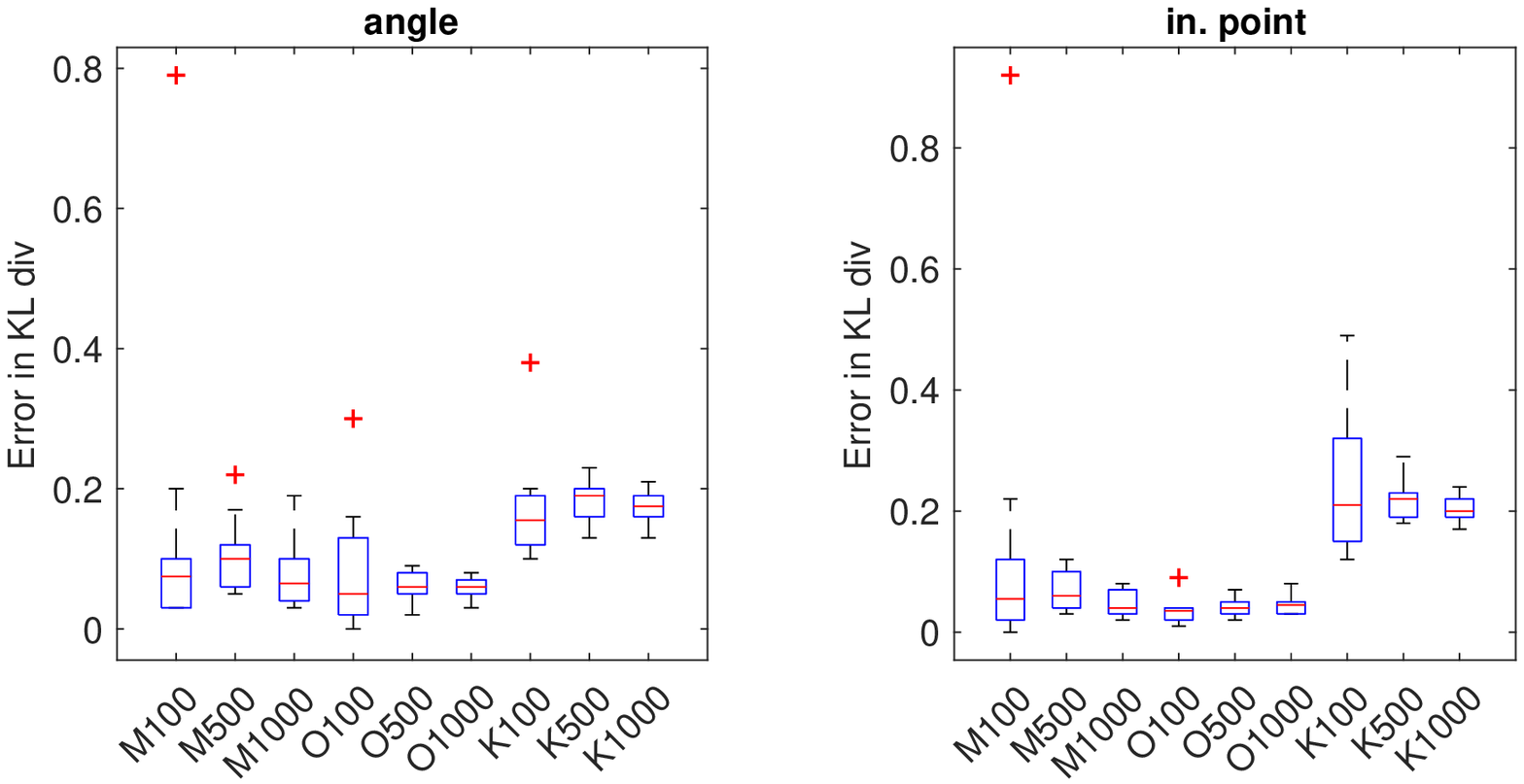}
  \includegraphics[width=0.9\textwidth]{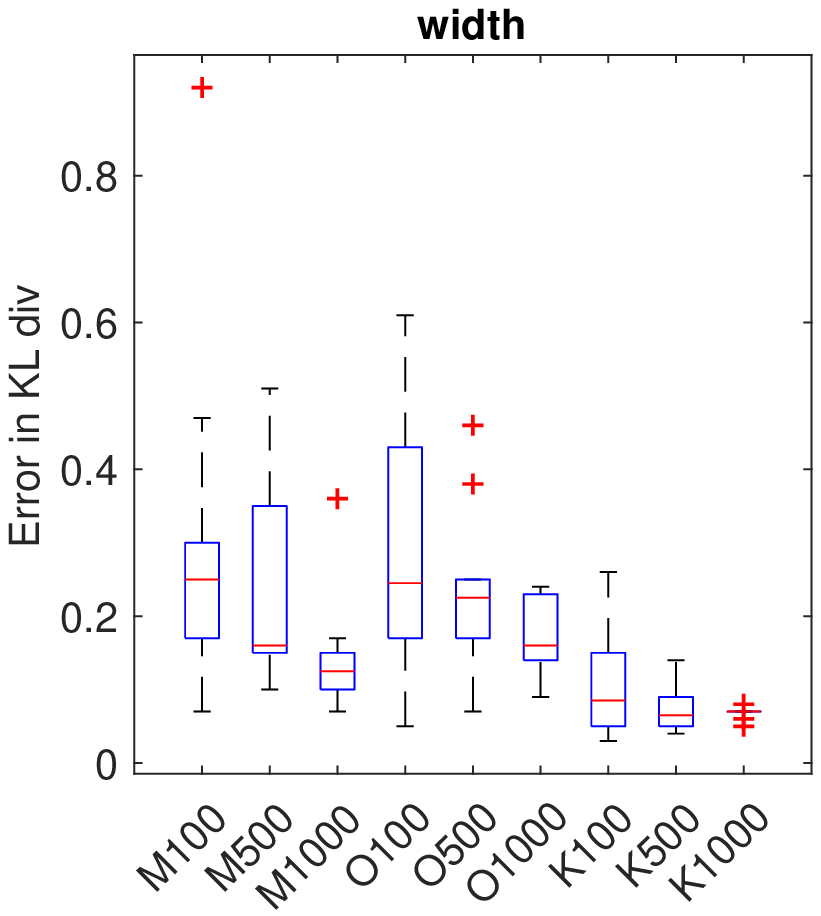}
  \caption{Box plot of KL divergence for geometrical parameters for P2 inference.  
 Central mark is the median, edges of the box are the
    25th and 75th percentiles, whiskers extend to the most extreme
    datapoints, and crosses are outliers.  
  On x-axis numbers stand for ensemble sizes, M stands for monomial based SMC, O for optimal transport based SMC, and K for EKI.  \label{fig:ChanKL}}
  \centering
\end{figure}

\begin{figure}[!ht]
  \centering
  \includegraphics[width=0.9\textwidth]{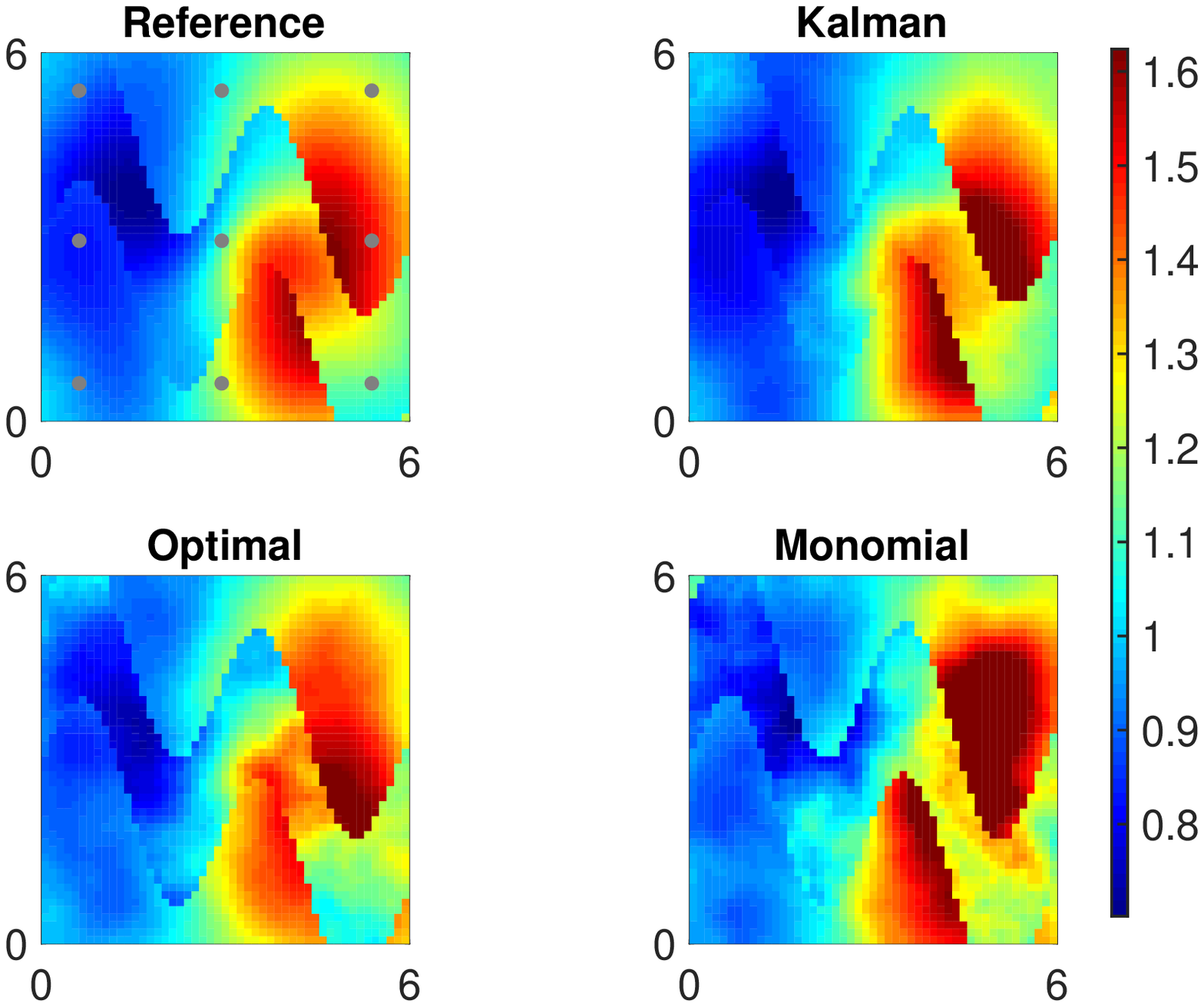}
  \caption{Mean permeability for P2 inference for the lowest error at ensemble size 1000. Observation locations are shown in circles.
\label{fig:ChanMF}}
  \centering
\end{figure}

In Figure~\ref{fig:ChanKL}, we plot KL divergence for geometrical parameters. We observe that EKI performs better than optimal transport based SMC for amplitude and width, while worse for other parameters. 
We should note that the two different modes of frequency shown in Figure 7 provide two significantly different channel configuration, thus it is important to correctly estimate the pdf.
Monomial based SMC performs comparably to 
optimal transport based SMC though not consistently better or worse. We should recall, however, that optimal transport based SMC outperforms monomial based SMC for log-permeability both inside and outside channel. 
In Figure~\ref{fig:ChanMF}, we show mean field of permeability over the channelized domain for the lowest error at ensemble size 1000.

\section{Conclusions}\label{sec:conc}
Accurate estimation of the posterior distribution of uncertain model parameters of strongly nonlinear problems remains a challenging problem. Parameters are high dimensional, they are not observed, and they do not have a dynamical equation. Moreover, due to nonlinearity of models even Gaussian prior of parameters might result in non-Gaussian posterior. 
Since MCMC is computationally unfeasible for high-dimensional problems, adaptive SMC is an alternative to estimate posterior distributions in the Bayesian framework. However, adaptive SMC still requires large ensembles. 

In order to reduce computational cost, we proposed to introduce optimal transport based resampling from~\cite{Re13} to adaptive SMC. 
Optimal transport based resampling creates new samples by maximizing variance between prior and posterior. 
It has been already shown for state estimation and parameter estimation with low dimension, that 
particle filter with optimal transport based resampling outperforms particle filter with monomial based resampling. 
As it was aimed to estimate time-evolving model states of chaotic systems, simple inflation was sufficient to mutate particles. 

Here we have adopted optimal transportation to elliptic Bayesian inverse problems. 
We have shown that optimal transport-based SMC has a high potential for Bayesian inversion of high-dimensional parameters. \mi{The parameterisation of the channelised permeability was particularly useful since it involves geometric parameters with marginal posteriors that display non-Gaussian features (e.g. bimodality in the frequency parameter; see Figure \ref{fig:ChanPDF}) which are often difficult to characterise via EKI. Indeed, for this case the proposed approach provides more accurate approximations to the marginal posteriors (quantified via KL divergence) than those approximated with EKI. Compared to the standard monomial-based SMC we did not observe substantial differences in the level of approximation of the aforementioned marginals. However, the proposed transport-based SMC outperforms the monomial-based version in approximating the high-dimensional (marginal) posteriors of the two spatially-variable log-permeability fields that we infer in the present setting (measured in terms of the error in the mean error and variance).

Moreover, optimal transport based SMC still underestimates variance (not shown), which could be improved by
considering second order consistent optimal transport resampling instead of first order.
However, second order consistent optimal transport resampling does not necessary provide with non-negative transformations.
Finally, optimal transport resampling does not need to be restricted to finite dimensions, at least theoretically~\cite{ChRe15}, with the challenge of finding such a minimizer computationally.}


\ack
This work is part of the research programme Shell-NWO/FOM Computational Sciences for Energy Research (CSER) with project number 14CSER007 which is partly financed by the Netherlands Organization for Scientific Research (NWO).

\bibliographystyle{iopart-num} 
\bibliography{RuDuIg18}

\end{document}